\documentclass[preprint]{aastex631}
\usepackage{hyperref}
\usepackage{url}
\usepackage{amsmath}
\usepackage{graphicx}
\usepackage{cleveref}
\usepackage{placeins}
\newcommand{\best}[1]{\textbf{#1}}

\shorttitle{Diffusion model for SN-Ia spectra}
\shortauthors{Shen and Gagliano}
\graphicspath{{./}{Figs/}}

\begin{document}

\title{Variational diffusion transformers for conditional sampling of supernovae spectra}

\author[0000-0003-2779-6507]{Yunyi Shen}
\affiliation{Massachusetts Institute of Technology \\
77 Massachusetts Avenue \\
Cambridge, MA, USA\\
\texttt{yshen99@mit.edu}
}

\author{Alexander T. Gagliano}
\affiliation{IAIFI, MIT\\
CfA, Harvard University \\
77 Massachusetts Avenue \\
Cambridge, MA, USA\\
\texttt{gaglian2@mit.edu} 
}

\begin{abstract}
Type Ia Supernovae (SNe~Ia) have become the most precise distance indicators in astrophysics due to their incredible observational homogeneity. Increasing discovery rates, however, have revealed multiple sub-populations with spectroscopic properties that are both diverse and difficult to interpret using existing physical models. These peculiar events are hard to identify from sparsely sampled observations and can introduce systematics in cosmological analyses if not flagged early; they are also of broader importance for building a cohesive understanding of thermonuclear explosions. In this work, we introduce \texttt{DiTSNe-Ia}, a variational diffusion-based generative model conditioned on light curve observations and trained to reproduce the observed spectral diversity of SNe~Ia. In experiments with realistic light curves and spectra from radiative transfer simulations, \texttt{DiTSNe-Ia} achieves significantly more accurate reconstructions than the widely used SALT3 templates across a broad range of observation phases (from 10 days before peak light to 30 days after it). \texttt{DiTSNe-Ia} yields a mean squared error of 0.108 across all phases—five times lower than SALT3’s 0.508—and an after-peak error of just 0.0191, an order of magnitude smaller than SALT3’s 0.305. Additionally, our model produces well-calibrated credible intervals with near-nominal coverage, particularly at post-peak phases. \texttt{DiTSNe-Ia} is a powerful tool for rapidly inferring the spectral properties of SNe~Ia and other transient astrophysical phenomena for which a physical description does not yet exist.
\end{abstract}


\section{Introduction}
\textbf{Spectral diversity of Type Ia Supernovae.} Type Ia supernovae (SNe~Ia) are the thermonuclear explosions of white dwarfs in binary systems as they approach the Chandrasekhar limit \citep{hoyle1960nucleosynthesis, liu2023type}. Since their use in confirming the accelerated expansion of the universe \citep{riess1998observational,perlmutter1999measurements}, SNe~Ia have been widely adopted as cosmological distance indicators. SNe~Ia are \textit{standardizable} candles: they exhibit correlations in their observational properties that can be corrected to infer their distance \citep[such as the ``Phillips relation" between an explosion's maximum luminosity and its dimming rate; ][]{pskovskii1977light, phillips1993absolute, phillips1999reddening}. 

Growing sample sizes of SNe~Ia have complicated the story. Wide-field imaging surveys such as the Zwicky Transient Facility \citep{bellm2018zwicky} have discovered otherwise-typical SNe~Ia with additional observational oddities (e.g., early and late-time re-brightening episodes), along with multiple sub-populations that defy the expected observational correlations. These include both red and underluminous SNe~Ia \citep[named ``SNe~Ia 1991bg-like" after their archetypal event; as well as SNe~Iax, alternatively known as ``SNe~2002cx-like";][]{filippenko1992subluminous}; and overluminous SNe~Ia \citep[SNe~Ia 1991T-like events;][]{filippenko1992peculiar}. Spectra of these SNe, which measure their wavelength-specific emission, have revealed important differences between the temperatures, velocities, and compositions of these more peculiar explosions. Nonetheless, our growing understanding of this observational diversity has only raised more questions about the physics of these explosions \citep{soker2019supernovae, liu2023type}.

\textbf{Legacy Survey of Space and Time.} The upcoming Vera C. Rubin Observatory Legacy Survey of Space and Time \citep[LSST,][]{ivezic2019lsst}, which will image the full Southern Sky every 3-4 days for ten years beginning in 2025, is expected to discover $\sim$1~M SNe each year \citep{ivezic2019lsst}. Half of them will be SNe~Ia. While discovery rates of these explosions will break exponential scaling as a result of the survey, our spectroscopic resources will only be able to observe the smallest subset (at most $\sim$0.1\%). Rapid and unbiased predictions of the spectroscopic behavior of discovered SNe~Ia from measurements of their time-evolving brightness (light curves) will allow astronomers to rapidly prioritize the most scientifically valuable events for spectroscopic observation. These data, in turn, will improve our ability both to precisely standardize the most homogeneous SNe~Ia for cosmology and to launch exploratory studies into the explosion properties of the most diverse. 

\textbf{Goal: spectra inference and generation.} Motivated by the need for low-latency physical inference in SN~Ia science, we consider the task of inferring spectra from light curves, with both obtained at optical wavelengths. These measurements are correlated slices of an event's spectral energy distribution (SED): light curves measure the evolution of an explosion's flux at specific wavelengths over time, while spectra measure flux changes in wavelength at a single moment in time \citep{zhang2024maven}. As a result, one can infer an explosion's spectral properties in time by sampling from a posterior of SN~Ia spectra conditioned on its light curve and time. This casts the inference problem as a conditional generation task. 

\section{A model for conditional spectra generation}
\begin{figure}[htp]
    \centering
    \vspace{-0.3cm}
    \includegraphics[width=0.8\linewidth]{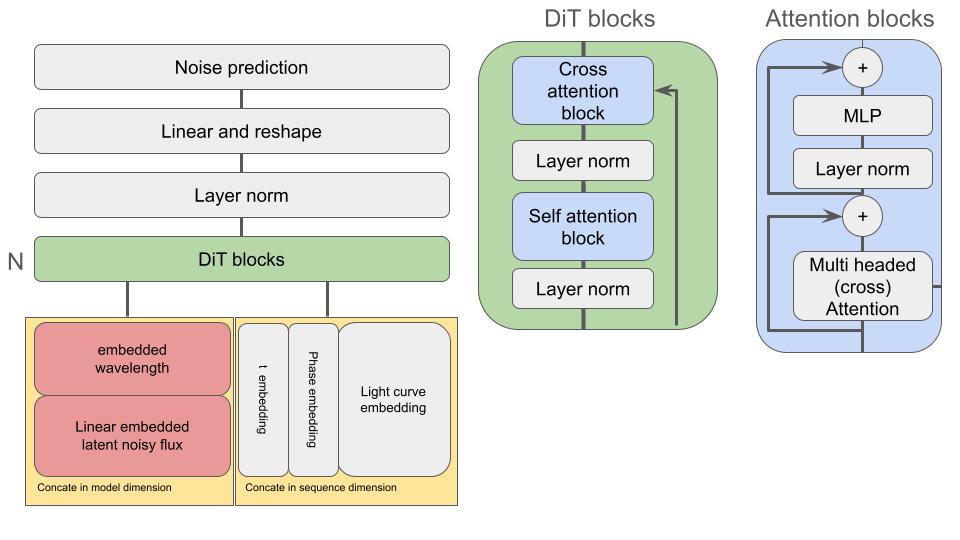}\vspace{-0.4cm}
    \caption{The architecture of our diffusion transformer for generation of SN~Ia spectra conditioned on light curve data. Positional embeddings are used to preserve correlations from data irregularly sampled in both time and wavelength.}
    \label{fig:architecture}
\end{figure}

\textbf{Variational diffusion transformer for spectra.}
We used a variational diffusion model for spectral generation \citep{kingma2021variational}. We used a transformer (DiT) architecture for the score model \citep{peebles2023scalable} because of the flexibility of modeling sequences that are not from a regular grid such as light curves and spectra. Each DiT block consists of one attention block and one cross-attention block separated by one layer normalization block. Each (cross)attention block also consists of multi-headed (cross)attention, layer normalization, and a Multi-Layer Perceptron (MLP), as well as skip connections (Figure~\ref{fig:architecture}). 

An ongoing challenge in training neural networks from astronomical observations is that the sequence may not be regularly-spaced over time and wavelength. Our model addresses this challenge by embedding wavelength/time (``positional") information first with a set of fixed sinusoidal functions, where we employ exponentially-changing frequencies as in the original transformer model \citep{vaswani2017attention} taking input of (normalized) wavelengths/times. 

We start with a sinusoidal embedding of dimension $2D_{\sin}$; $P$ denotes the sinusoidal encoding of positional data (wavelength and time), with $s\in \mathcal{S}$ for dimension $2i,2i+1$ and $i=0,\dots, D_{\sin}$. The final positional embedding $\bm{E}$ is then given by
\begin{equation}
    \begin{aligned}
    &P_{2i}(s)= \sin\left(\frac{s}{10,000^{2i/D_{\sin}}}\right)\quad P_{2i+1}(s) = \cos\left(\frac{s}{10,000^{2i/D_{\sin}}}\right)\\
    &\bm{E}(s)=MLP(\bm{P}(s))
\end{aligned}
\label{eq:positiona}
\end{equation}
We then pass the encoding of size $[L_{spec}, 2D_{\sin}]$, with $L_{spec}$ being the length of spectra, through an MLP to reshape it into the model dimension $D$. This is a similar treatment as diffusion time in \citet{peebles2023scalable}.

For our spectral observations, we embed wavelength data with the above treatment, add it to a linear embedding of flux, and project to the model dimension. This will yield an internal representation of size $[L_{spec}, D]$ with $D$ being the dimension of the model and $L_{spec}$ being the length of the spectrum. We then pass it through diffusion transformer blocks to predict the added noise. We discuss our treatment of light curve observations below. 

\textbf{Conditioning and photometric embedding.} In our conditional generation framework, there are three datasets to condition on: the diffusion time of the model, the phase of the observation (the time relative to the supernova's maximum light, in days), and the light curve observations themselves. We adopt a cross-attention strategy in our model due to the correlated information between these data and the complexity of the unstructured light curve sequences. Diffusion time and phase were first passed through the fixed sinusoidal functions (Figure~\ref{eq:positiona}) and then projected to the model dimension using an MLP with learnable weights. Each of them yield a representation of size $[1, D]$ for model dimension $D$. 

Light curve data were embedded similar to spectra: we first embedded phase using fixed sinusoidal functions, passed the output to a learnable MLP, and then added these to a linear embedding of the brightness measurements in magnitudes. To the combined vector, we also add an embedding of the transmission filter of the observation. These transmission filters were treated as categorical variables and the embeddings are simple look-up tables. This filter information encodes the range of wavelengths at which the brightness of the supernova was measured. This will yield a sequence embedding for the light curve of size $[L_{LC},D]$, with $L_{LC}$ being the sequence length of the light curves and $D$ being the common model dimension size. 

Representations of light curve, phase and diffusion time are then concatenated in the sequence dimension to get a conditional representation of size $[L_{LC}+2, D]$. 

We show an overview of the \texttt{DiTSNe-Ia} architecture in Figure~\ref{fig:architecture}. The model has a dimension $D=256$. The MLP used in each DiT block consists of a single hidden layer with dimension 512. In both multi-headed attention and cross-attention, we use 4 heads and 6 layers of DiT blocks.

\textbf{Loss and training.} We train our model to minimize the variational lower bound (VLB) on the marginal likelihood \citep[as in Equation 11 of ][]{kingma2021variational}. 

\section{Experiments}
\textbf{Data, preprocessing, and training scheme.}
We use the model grid of \citet{goldstein2018evidence} to train and validate \texttt{DiTSNe-Ia}. \citet{goldstein2018evidence} ran 4,500 simulations of SNe~Ia with the radiation transport code \texttt{Sedona} \citep{kasen2006secondary} to produce a full SED surface in temporal bins of 1~day and wavelength bins of ~30~Å. The simulated SNe~Ia span a broad range of ejecta masses $\left(M_{ej}\in[0.7\;M_{\odot}, 2.5\;M_{\odot}]\right)$, velocities $\left(v_K\in[8000\;\mathrm{km}\;\mathrm{s}^{-1},15000\;\mathrm{km}\;\mathrm{s}^{-1}]\right)$, masses of the exploding star $\left(m\in[1\;M_{\odot},2.5\;M_{\odot}]\right)$ (where the traditional Chandrasekhar limit corresponding to a thermonuclear explosion is $\approx1.4\;M_{\odot}$); and fractions of the ejecta composed of both radioactive $^{56}\mathrm{Ni}$ $\left(f_{\mathrm{Ni}}\in[0.1,0.8]\right)$ and C/O $\left(f_{\mathrm{CO}}\in[0.00, 0.07]\right)$. We split the sample into 80\% training, 20\% testing. We then generated LSST-like light curves by integrating the SED surfaces through the published transmission curves of the LSST filters from the SVO Filter Profile Service\footnote{\url{http://svo2.cab.inta-csic.es/}}. We model the light curve cadence using the baseline v3.3 simulations of LSST's Wide-Fast-Deep survey and assume a random distribution of events in space across the survey footprint and in time since survey start. We have assumed that all SNe occur at a distance of 10~pc; in future work, we plan to adopt a SN~Ia volumetric rate to evaluate the robustness of the model on light curves with realistic noise properties. 

We impose quality cuts on our generated light curves, requiring at least 10 total measurements across all filters and with observations in any band before and after the explosion's peak brightness. For our simulated spectra, we calculate the flux directly from the SED surface in 10-day windows from 10 days before peak light to 30 days following peak light; we apply a median filter with bandwidth 3 to these data to mimic the finite resolution of a spectrograph. We then train the model using the log of the flux values in units of $\mathrm{erg}\;\mathrm{s}^{-1}\;\mathrm{cm}^{-2}\;\mathrm{Hz}^{-1}$. 

We standardized all flux, wavelength, phase, and magnitude values to z-scores for training, and convert them back to physical units at inference time. We also pad the light curves to 10 measurements in each band, resulting in uniform light curve arrays of length 60. 

We train \texttt{DiTSNe-Ia} in \texttt{jax} \citep{jax2018github} using the AdamW optimizer \citep{loshchilov2017fixing}. We used a warm-up cosine decay schedule as the learning rate with 500 warm-up steps and 3000 decay steps, a maximum learning rate of $3\times 10^{-3}$, an initial learning rate of $10^{-5}$, and a batch size of 64. 

We tested our method and the baseline model on 300 SNe from the test set at five phases (-10 to 30 days relative to peak light in 10-day bins). For each light curve and phase combination, we took 100 samples of each conditional generation, used averages as point estimates, and calculate equal-tail credible intervals of each prediction (CI).

\textbf{Test metric}
We used three metrics to test the proposed method, one for prediction accuracy and two for uncertainty quantification 
\begin{itemize}
    \item \textbf{Residual} of point estimate: the difference between point estimate of the model and ground truth spectra at different days after peak. We also quantify the accuracy using the mean squared error (MSE).
    \item \textbf{Coverage}: the empirical probability that the posterior CI contains the ground truth value as a function of wavelength.
    \item \textbf{CI width}: the difference between the upper quantile and lower quantile of the CI for a given level.
\end{itemize}

\textbf{Baseline.}
We compare our prediction to SALT3 \citep{kenworthy2021salt3, guy2007salt2}, a popular template-matching technique for SN~Ia spectra. The SALT3 model consists of linear combinations of SN~Ia SED components, whose contributions are determined by parameters ($x_0$, $x_1$, and $c$) fit to light curves in multiple filters. We used the implementation of the SALT3 surfaces in the python package \texttt{sncosmo} \citep{barbary2016sncosmo} and fit the surface to our re-sampled light curves with an MCMC to quantify fit uncertainties. 

\begin{figure}[htp]
    \centering
    \vspace{-0.25cm}
    \includegraphics[width=\linewidth]{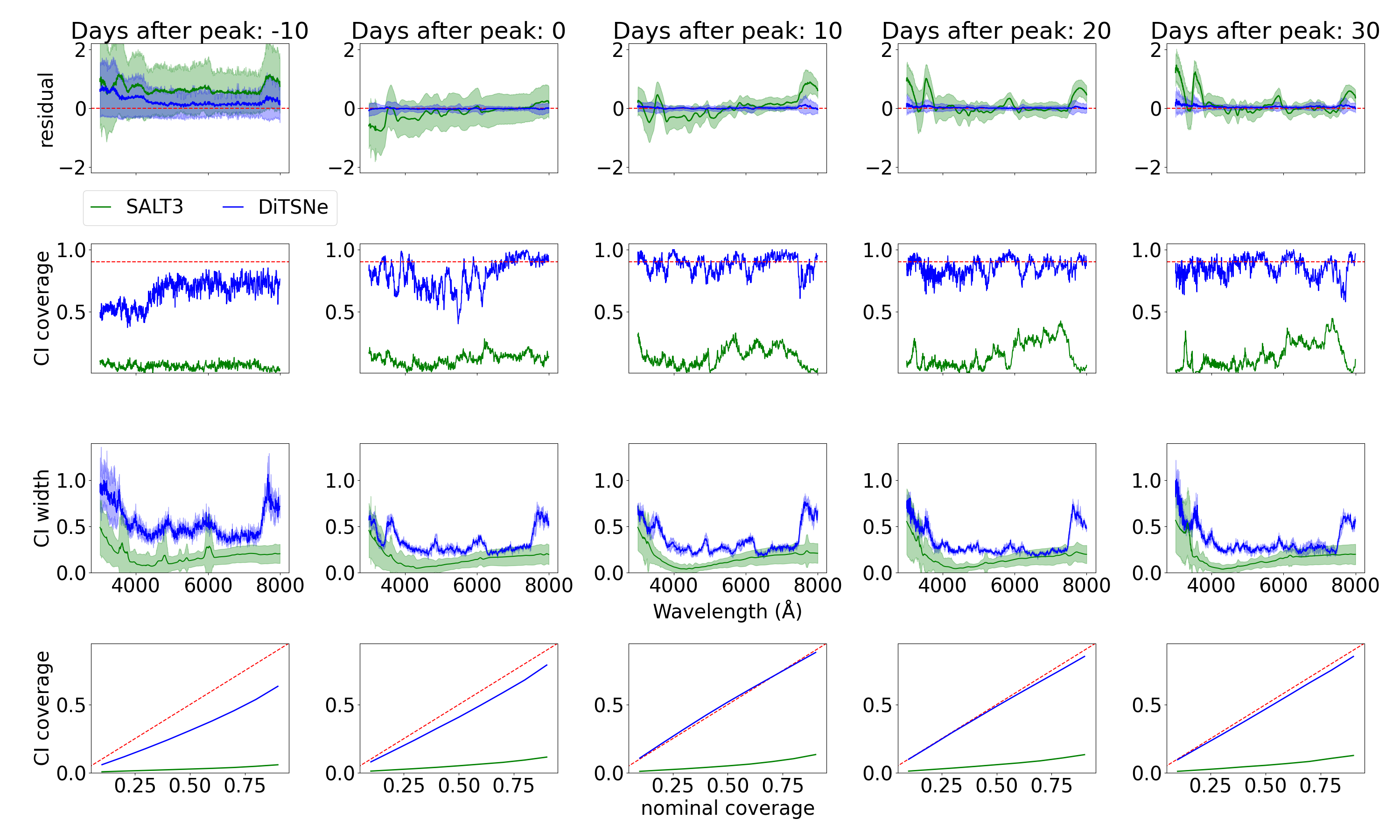}
    \vspace{-0.5cm}
    \caption{\texttt{DiTSNe} Prediction - Ground Truth and associated 1$\sigma$ uncertainties (First Row) for spectra from the test set predicted over five different phases and conditioned on realistic LSST light curves. The prediction of the baseline SALT3 model is shown in green. Our method achieves a lower mean error rate than SALT3 at all phases. We also show coverage plots (Second Row) from the model, finding values near nominal starting +10 days after peak light; as well as the width of the width of the 90\% CI from \texttt{DiTSNe} predictions (Third Row) and its realized coverage as a function of nominal coverage (Fourth Row).}
    \label{fig:prediction}
\end{figure}

\textbf{Results.}
We provide a summary of our results in \Cref{tab:summaryresults}, and in \Cref{fig:prediction}. We find that \texttt{DiTSNe-Ia} achieves better predictive performance across the test set while having a lower average bias than the SALT3 method. \texttt{DiTSNe-Ia} yields an overall mean squared error of 0.108—five times lower than SALT3’s 0.508—and an after-peak error of just 0.0191, an order of magnitude smaller than SALT3’s 0.305. Uncertainty quantification achieves coverage near nominal values in predictions greater than 10 days after peak light with an overall coverage of 0.845 for 90\% interval (see also last row in Figure~\ref{fig:prediction}); although the predictions at earlier phases (especially 10 days before peak) are under-covered. This is unsurprising given the limited photometric information available at early times. SALT3, in contrast, has small CIs but is severely under-covered. 

\begin{figure}[htp]
    \centering
    \vspace{-0.25cm}
    \includegraphics[width=\linewidth]{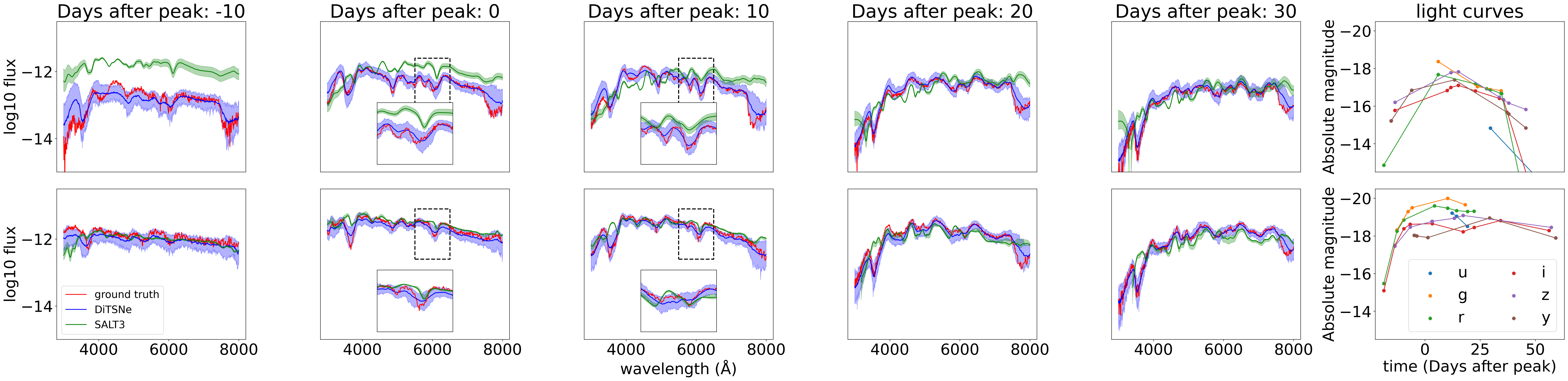}
    \includegraphics[width=\linewidth]{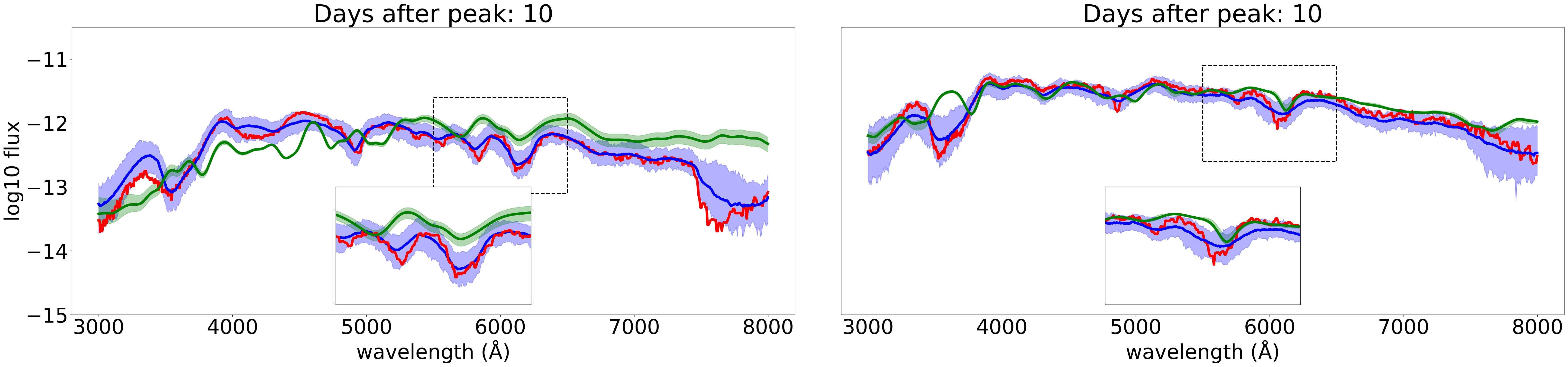}
    \vspace{-0.25cm}
    \caption{Upper: Spectra and light curves for two randomly chosen SNe~Ia from the test set, with zoom-in of finer-scale features. The flux was calculated in units of $\text{erg s}^{-1}\text{cm}^{-2}\;\text{Hz}^{-1}$ and light curves are in absolute magnitude. Lower: Zoom-in view of each model's predictions 10 days after peak. Our method can recover finer-scale spectral features associated with the physical properties of the explosion that are not well-captured by SALT3.}
    \label{fig:example}
\end{figure}

\Cref{fig:example} shows the spectra posteriors for two randomly-selected SNe~Ia in the test set as a function of phase. Qualitatively, we find that our method not only recovers the broad shape of each spectrum (which encodes information about the temperature and explosion energy of the SN); but also finer-scale features (which encode the explosion's composition and velocity structure) more precisely than the baseline model. We plan to quantify the constraints on these physical properties in a future study.

\begin{table}[htp]
    \centering
    {
    \small
    \begin{tabular}{llccccc|cc}
    \hline
      Metric &Method  & -10 days & 0 days & 10 days & 20 days & 30 days & Mean & After Peak\\
      \hline
       MSE &SALT3 & 1.49& 0.467& 0.232& 0.247& 0.279& 0.508& 0.305 \\ 
        & DiTSNe-Ia & \best{0.465} & \best{0.0183} & \best{0.015} & \best{0.0164} & \best{0.0267}& \best{0.108}& \best{0.0191} \\
        \hline
        CI Coverage & SALT3  &0.0639 & 0.115& 0.134& 0.137& 0.131 & 0.116& 0.129 \\
        & DiTSNe-Ia  &\best{0.634} & \best{0.79} & \best{0.883} & \best{0.854} & \best{0.853} & \best{0.803}& \best{0.845}\\
       \hline
       CI Width & SALT3 &\best{0.187}& \best{0.154}& \best{0.157}& \best{0.165}& \best{0.174} & \best{0.165} & \best{0.161}\\
         &DiTSNe-Ia &0.548& 0.332& 0.353& 0.336& 0.369 & 0.387& 0.347 \\
    \hline
    \end{tabular}
    }
    \caption{
    Average MSE, CI coverage, and CI widths over all SNe in the test set as a function of phase (number of days relative to the peak brightness of each SN). Our method shows better performance than the SALT3 baseline but can be under-covered. Coverage and widths are shown for the 90\% CI of predicted \texttt{DiTSNE-Ia} spectra. SALT3, in contrast, has shorter CIs but is severely under-covered.}
    \label{tab:summaryresults}
\end{table}

\section{Discussion}
We have presented \texttt{DiTSNe-Ia}, a state-of-the-art generative model for inferring SN Ia spectra at multiple phases during an explosion using light curve information under idealized observing conditions. The method performs well when applied to simulated SNe Ia from \citet{goldstein2018evidence}, demonstrating the potential of diffusion-based generative models in SN science.

Our follow-up work will focus on connecting the model's performance, measured by MSE and CI coverage, to physical parameters in the simulation to identify systematic patterns. Several challenges remain. For instance, although our approach is motivated by observational constraints, the current training data are still based on simulations, potentially producing a model that underestimates the true diversity observed in real data. Additionally, we assume that the training light curves and spectra are corrected for observational effects such as redshift. However, in practice, redshift is not always fully known and carries its own uncertainties.

Another area for improvement is the representation of light curves, particularly in incorporating translation invariance—a property useful when the exact time of SN peak is unknown. Relative positional encoding \citep{shaw2018self} does not directly generalize to cross-attention mechanisms, as it requires both sequences to exist in the same space with well-defined positional distances, which is not the case for light curves and spectra. A promising direction may be to apply convolutional layers before the attention mechanism to extract translation-invariant features, a strategy commonly used in the computer vision community \citep{wu2021cvt,xiao2021early}. We will investigate these improvements to the model in future work.


\section{Software and third party data repository citations} \label{sec:cite}
Astropy, sncosmo, jax, numpy

\section*{Acknowledgements}
\begin{acknowledgments}
This work is supported by the National Science Foundation under Cooperative Agreement PHY-2019786 (The NSF AI Institute for Artificial Intelligence and Fundamental Interactions, http://iaifi.org/
\end{acknowledgments}

%








\bibliography{references}{}

\begin{thebibliography}{}
\expandafter\ifx\csname natexlab\endcsname\relax\def\natexlab#1{#1}\fi
\providecommand{\url}[1]{\href{#1}{#1}}
\providecommand{\dodoi}[1]{doi:~\href{http://doi.org/#1}{\nolinkurl{#1}}}
\providecommand{\doeprint}[1]{\href{http://ascl.net/#1}{\nolinkurl{http://ascl.net/#1}}}
\providecommand{\doarXiv}[1]{\href{https://arxiv.org/abs/#1}{\nolinkurl{https://arxiv.org/abs/#1}}}

\bibitem[{Barbary {et~al.}(2016)Barbary, Barclay, Biswas, Craig, Feindt,
  Friesen, Goldstein, Jha, Rodney, Sofiatti, {et~al.}}]{barbary2016sncosmo}
Barbary, K., Barclay, T., Biswas, R., {et~al.} 2016, Astrophysics Source Code
  Library, ascl

\bibitem[{Bellm {et~al.}(2018)Bellm, Kulkarni, Graham, Dekany, Smith, Riddle,
  Masci, Helou, Prince, Adams, {et~al.}}]{bellm2018zwicky}
Bellm, E.~C., Kulkarni, S.~R., Graham, M.~J., {et~al.} 2018, Publications of
  the Astronomical Society of the Pacific, 131, 018002

\bibitem[{Bradbury {et~al.}(2018)Bradbury, Frostig, Hawkins, Johnson, Leary,
  Maclaurin, Necula, Paszke, Vander{P}las, Wanderman-{M}ilne, \&
  Zhang}]{jax2018github}
Bradbury, J., Frostig, R., Hawkins, P., {et~al.} 2018, {JAX}: composable
  transformations of {P}ython+{N}um{P}y programs, 0.3.13.
\newblock \url{http://github.com/jax-ml/jax}

\bibitem[{Filippenko {et~al.}(1992{\natexlab{a}})Filippenko, Richmond, Branch,
  Gaskell, Herbst, Ford, Treffers, Matheson, Ho, Dey,
  {et~al.}}]{filippenko1992subluminous}
Filippenko, A.~V., Richmond, M.~W., Branch, D., {et~al.} 1992{\natexlab{a}},
  Astronomical Journal (ISSN 0004-6256), vol. 104, no. 4, p. 1543-1556, 1684.,
  104, 1543

\bibitem[{Filippenko {et~al.}(1992{\natexlab{b}})Filippenko, Richmond,
  Matheson, Shields, Burbidge, Cohen, Dickinson, Malkan, Nelson, Pietz,
  {et~al.}}]{filippenko1992peculiar}
Filippenko, A.~V., Richmond, M.~W., Matheson, T., {et~al.} 1992{\natexlab{b}},
  Astrophysical Journal, Part 2-Letters (ISSN 0004-637X), vol. 384, Jan. 1,
  1992, p. L15-L18., 384, L15

\bibitem[{Goldstein \& Kasen(2018)}]{goldstein2018evidence}
Goldstein, D.~A., \& Kasen, D. 2018, The Astrophysical Journal Letters, 852,
  L33

\bibitem[{Guy {et~al.}(2007)Guy, Astier, Baumont, Hardin, Pain, Regnault, Basa,
  Carlberg, Conley, Fabbro, {et~al.}}]{guy2007salt2}
Guy, J., Astier, P., Baumont, S., {et~al.} 2007, Astronomy \& Astrophysics,
  466, 11

\bibitem[{Hoyle \& Fowler(1960)}]{hoyle1960nucleosynthesis}
Hoyle, F., \& Fowler, W.~A. 1960, Astrophysical Journal, vol. 132, p. 565, 132,
  565

\bibitem[{Ivezi{\'c} {et~al.}(2019)Ivezi{\'c}, Kahn, Tyson, Abel, Acosta,
  Allsman, Alonso, AlSayyad, Anderson, Andrew, {et~al.}}]{ivezic2019lsst}
Ivezi{\'c}, {\v{Z}}., Kahn, S.~M., Tyson, J.~A., {et~al.} 2019, The
  Astrophysical Journal, 873, 111

\bibitem[{Kasen(2006)}]{kasen2006secondary}
Kasen, D. 2006, The Astrophysical Journal, 649, 939

\bibitem[{Kenworthy {et~al.}(2021)Kenworthy, Jones, Dai, Kessler, Scolnic,
  Brout, Siebert, Pierel, Dettman, Dimitriadis, {et~al.}}]{kenworthy2021salt3}
Kenworthy, W., Jones, D., Dai, M., {et~al.} 2021, The Astrophysical Journal,
  923, 265

\bibitem[{Kingma {et~al.}(2021)Kingma, Salimans, Poole, \&
  Ho}]{kingma2021variational}
Kingma, D., Salimans, T., Poole, B., \& Ho, J. 2021, Advances in neural
  information processing systems, 34, 21696

\bibitem[{Liu {et~al.}(2023)Liu, R{\"o}pke, \& Han}]{liu2023type}
Liu, Z.-W., R{\"o}pke, F.~K., \& Han, Z. 2023, Research in Astronomy and
  Astrophysics, 23, 082001

\bibitem[{Loshchilov {et~al.}(2017)Loshchilov, Hutter,
  {et~al.}}]{loshchilov2017fixing}
Loshchilov, I., Hutter, F., {et~al.} 2017, arXiv preprint arXiv:1711.05101, 5

\bibitem[{Peebles \& Xie(2023)}]{peebles2023scalable}
Peebles, W., \& Xie, S. 2023, in Proceedings of the IEEE/CVF International
  Conference on Computer Vision, 4195--4205

\bibitem[{Perlmutter {et~al.}(1999)Perlmutter, Aldering, Goldhaber, Knop,
  Nugent, Castro, Deustua, Fabbro, Goobar, Groom,
  {et~al.}}]{perlmutter1999measurements}
Perlmutter, S., Aldering, G., Goldhaber, G., {et~al.} 1999, The Astrophysical
  Journal, 517, 565

\bibitem[{Phillips {et~al.}(1999)Phillips, Lira, Suntzeff, Schommer, Hamuy, \&
  Maza}]{phillips1999reddening}
Phillips, M., Lira, P., Suntzeff, N.~B., {et~al.} 1999, The Astronomical
  Journal, 118, 1766

\bibitem[{Phillips(1993)}]{phillips1993absolute}
Phillips, M.~M. 1993, Astrophysical Journal, Part 2-Letters (ISSN 0004-637X),
  vol. 413, no. 2, p. L105-L108., 413, L105

\bibitem[{Pskovskii(1977)}]{pskovskii1977light}
Pskovskii, I.~P. 1977, Soviet Astronomy, vol. 21, Nov.-Dec. 1977, p. 675-682.
  Translation. Astronomicheskii Zhurnal, vol. 54, Nov.-Dec. 1977, p.
  1188-1201., 21, 675

\bibitem[{Riess {et~al.}(1998)Riess, Filippenko, Challis, Clocchiatti, Diercks,
  Garnavich, Gilliland, Hogan, Jha, Kirshner,
  {et~al.}}]{riess1998observational}
Riess, A.~G., Filippenko, A.~V., Challis, P., {et~al.} 1998, The astronomical
  journal, 116, 1009

\bibitem[{Shaw {et~al.}(2018)Shaw, Uszkoreit, \& Vaswani}]{shaw2018self}
Shaw, P., Uszkoreit, J., \& Vaswani, A. 2018, arXiv preprint arXiv:1803.02155

\bibitem[{Soker(2019)}]{soker2019supernovae}
Soker, N. 2019, New Astronomy Reviews, 87, 101535

\bibitem[{Vaswani(2017)}]{vaswani2017attention}
Vaswani, A. 2017, Advances in Neural Information Processing Systems

\bibitem[{Wu {et~al.}(2021)Wu, Xiao, Codella, Liu, Dai, Yuan, \&
  Zhang}]{wu2021cvt}
Wu, H., Xiao, B., Codella, N., {et~al.} 2021, in Proceedings of the IEEE/CVF
  international conference on computer vision, 22--31

\bibitem[{Xiao {et~al.}(2021)Xiao, Singh, Mintun, Darrell, Doll{\'a}r, \&
  Girshick}]{xiao2021early}
Xiao, T., Singh, M., Mintun, E., {et~al.} 2021, Advances in neural information
  processing systems, 34, 30392

\bibitem[{Zhang {et~al.}(2024)Zhang, Helfer, Gagliano, Mishra-Sharma, \&
  Villar}]{zhang2024maven}
Zhang, G., Helfer, T., Gagliano, A.~T., Mishra-Sharma, S., \& Villar, V.~A.
  2024, Machine Learning: Science and Technology, 5, 045069

\end{thebibliography}
\bibliographystyle{aasjournal}



\end{document}